\documentclass[sigconf]{acmart}

\usepackage{booktabs} % For formal tables

\setcopyright{rightsretained}
\usepackage{multirow}
\usepackage{rotating}
\usepackage{latexsym}
\usepackage{pifont}
\newcommand{\tick}{\ding{52}}

\usepackage{amsfonts}

\def\IE{{\textsf IE}}
\def\IE{{\textsf IE}}
\def\WIN{{\textsf WINDOWS}}
\def\PLU{{\textsf PLUGIN}}
\def\COM{{\textsf PROD}}

\def\SE[#1]{{$S_{Ex,{#1}}$}}
\def\SP[#1]{{$S_{PoC,{#1}}$}}
\def\SB[#1]{{$S_{Black,{#1}}$}}

\def\S3{{$S_{3rd}$}}

\def\NVD{{\textsf NVD}}

\def\SYM{{\textsf SYM}}
\def\WINE{{\textsf WINE}}

\begin{document}

\title{Attack Potential in Impact and Complexity
%\\An analysis of Symantec's Threats and BlackMarket Exploit Kits
}

\author{Luca~Allodi}
\affiliation{Eindhoven University of Technology}
\email{l.allodi@tue.nl}
\author{Fabio~Massacci} 
\affiliation{University of Trento}
\email{fabio.massacci@unitn.it}

\begin{abstract}
Vulnerability exploitation is reportedly one of the main attack
vectors against computer systems. Yet, most vulnerabilities remain unexploited by attackers. It is therefore of central importance to identify vulnerabilities that carry a high `potential for attack'.
In this paper we rely on Symantec data on real attacks detected in the wild to
identify a trade-off in the Impact and Complexity of a vulnerability, 
in terms of attacks that it generates; 
exploiting this effect, we devise a readily computable 
estimator of the vulnerability's \emph{Attack Potential} that
reliably estimates the expected volume of attacks against the vulnerability. 
We evaluate our estimator performance against standard patching
policies by measuring foiled attacks and demanded workload expressed as the number of vulnerabilities entailed to patch. We
show that our estimator significantly improves over standard patching policies by ruling out low-risk vulnerabilities, while maintaining invariant levels of coverage against attacks in the wild. Our estimator can be used as a first aid for
vulnerability prioritisation to focus assessment efforts on high-potential vulnerabilities.
\end{abstract}



% We no longer use \terms command


\copyrightyear{2017} 
\acmYear{2017} 
\setcopyright{acmlicensed}
\acmConference{ARES '17}{August 29-September 01, 2017}{Reggio Calabria, Italy}\acmPrice{15.00}\acmDOI{10.1145/3098954.3098965}
\acmISBN{978-1-4503-5257-4/17/08}

\maketitle

\section{Introduction}

% The recent increasing trend in automation of cyber-attacks 
% \cite{symantec-ekits-2011,Provos-2011-GOOGLE}, and the expanding
% importance of the underground economy in the general threat scenario
% \cite{Allodi-2013-IWCC,Grier-12-CCS}, leaves one wondering
% if such trends are also reflected in the nature of vulnerabilities that are exploited. 
The identification of objective and readily available measures for 
vulnerability risk is a central part of the vulnerability mitigation process~\cite{Verizon-2014,PCI-DSS-DOC,SCAR-MELL-09-ESEM}.
Industry standards such as the Common Vulnerability Scoring System (CVSS) have been developed to create a common framework over which evaluate vulnerability severity and guide the vulnerability mitigation process~\cite{SCAR-MELL-09-ESEM}; a CVSS assessment produces two final components: 
% \begin{itemize}

1) The \texttt{CVSS-vector}, that contains the fine-grained information regarding the characteristics of the vulnerability. Among other values, the \texttt{CVSS-vector} provides information on the \emph{complexity} of the vulnerability exploitation (in the metric \texttt{Access Complexity}), and its impact on the attacked system.

2) The \texttt{CVSS-score}, a final severity score that synthesises the information in the \texttt{CVSS-vector} in a single value between 0 and 10 (less severe to more severe).
%\end{itemize}

The \texttt{CVSS-score} is widely used as a metric for vulnerability management;
  for example, PCI-DSS, the worldwide security standard for systems handling credit-card data, sets a `hard threshold' for vulnerability patching at a CVSS score greater or equal to 4 (10 being the maximum)~\cite{PCI-DSS-DOC}. Similarly, NIST's SCAP standard uses the \texttt{CVSS-score} as the metric of reference for vulnerability assessment across industry sectors, including consumer systems~\cite{Scarfone-2010-SCAP}. 

Unfortunately, recent studies show that the \texttt{CVSS-score} does not correlate well with attacks in the wild, leading to sub-optimal vulnerability management policies~\cite{Allodi-2014-TISSEC,Christey-2013-BHUSA,Verizon-2014}.
% This poses an interesting question in identifying which
% vulnerabilities the attacker would choose exploit.
% Previous
% work showed that attackers seem to prefer certain
% vulnerabilities over others \cite{Allodi-ESSOS-15,Nayak-2014-RAID}.
% In the industry, the standard-de-facto approach to identify
% vulnerabilities at high risk of exploitation is the usage of the Common Vulnerability Scoring System, or CVSS in short, as a risk metric
% for vulnerabilities \cite{Mell-2007-CMU,Allodi-2014-TISSEC}.
% The CVSS score is widely used in scientific studies
% and international standards alike \cite{Scarfone-2010-SCAP}.
% Unfortunately, the CVSS scores has already been shown to be not a good proxy metric for risk of cyber-attacks \cite{BOZORGI-etal-10-SIGKDD,Allodi-2014-TISSEC}.
This is particularly unfortunate as the CVSS score gives a clear, well-defined 
and readily available assessment of the vulnerability that can be
used `out-of-the-box' to take a first security decision on whether the vulnerability is (not) likely to represent a significant risk~\cite{chakradeo2013mast}. This is especially relevant as recent empirical~\cite{Allodi-ESSOS-15} as well as analytical~\cite{Allodi-17-WAAM} findings indicate that most vulnerabilities remain unexploited by attackers. It is therefore especially important to devise measures that rule out `low-risk' vulnerabilities to prioritize fine-grained assessments on high-potential vulnerabilities.

%In this paper, we evaluate whether the CVSS 
%subscores may provide a more interesting estimation of the risk represented by a vulnerability. 
% The general CVSS score takes into consideration two subscores: \emph{Impact}
% and \emph{Exploitability}.
% The Impact subscore estimates the criticality of the vulnerability
% exploitation on the system;
% % and is
% % computed on the traditional
% % \textsf{Confidentiality, Integrity, Availability} (CIA) assessment.
% the Exploitability subscore is a measure of how likely the
% exploitation is
% \cite{BOZORGI-etal-10-SIGKDD}, and is computed on three assessments on
% \textsf{Access Vector, Authentication, \texttt{Access Complexity}}. In particular,
% in this manuscript
% we focus on the latter, which is an estimation of the difficulty of an attack
% targeting that vulnerability.
 
In this paper we investigate whether the information reported in the \texttt{CVSS-vector} may provide useful information, otherwise lost in the aggregate score, to estimate the attack potential of a vulnerability. %, and evaluate resulting patching policies against records of real attacks in the wild. 
% The \texttt{CVSS-vector} reports assessments on the \emph{Impact}
% and \emph{Exploitability} of the vulnerability.
% The Impact subscore estimates the criticality of the vulnerability
% exploitation on the system;
%  and is
%  computed on the traditional
%  \textsf{Confidentiality, Integrity, Availability} (CIA) assessment.
% the Exploitability subscore is a measure of how likely the
% exploitation is
% \cite{BOZORGI-etal-10-SIGKDD}, and is computed on three assessments on
% \textsf{Access Vector, Authentication, \texttt{Access Complexity}}. In particular,
% in this manuscript
% we focus on the latter, which is an estimation of the difficulty of successfully launching an attack against the vulnerability.
Leveraging on real attack data from Symantec, we show the existence of a clear 
trade-off between exploitation complexity and impact in terms of number of 
expected attacks observed in the wild. 
% By analysing trends of attacks per CIA assessment and complexity,
% %  we first show that the impact assessment
% % of a vulnerability can be greatly simplified with respect to current approaches
% % without losing insight detail. Second, 
% we show a clear correlation between number of 
% of the existence of a trade-off in vulnerability exploitation in terms
% of \emph{complexity} of the exploitation and \emph{impact} of the attack. 
% Based on this trade-off we identify
% `highly profitable' vulnerabilities and `low
% profitable' vulnerabilities. 
We propose to leverage this trade-off to identify a new measure, `\emph{Attack Potential}', that can be readily estimated from the \texttt{CVSS-vector} and used as a measure of vulnerability prioritization next to the standard \texttt{CVSS-score} metric (e.g. in a standard \emph{security triage} process~\cite{chakradeo2013mast}).
% Of course, if a vulnerability has High Impact
% and Low Complexity it does not mean that it will certainly be exploited
% millions of times in the wild. Rather, 
% Our estimator is an extension over current practices
% relying on the sole CVSS score: if the vulnerability should be fixed,
% an estimation of its potential exploitation in the wild can be useful
% in further defining the priority of the patching work.
% build an estimator
% of the potential volume of 
% attacks in the wild that will target a vulnerability. 
% This can be useful in particular when prioritising patching work to assign
% higher or lower patching priority to vulnerabilities that have a higher
% potential of being attacked. 
% To validate the estimator we
% rely on \WINE, a data-sharing initiative run by Symantec that gives
% access to researchers worldwide to real attack data. 
% Using a sample of vulnerabilities from \WINE\ we show
% that our estimator reliably estimates the attack potential of a vulnerability
% without `undershooting', i.e. estimating a low potential
%  of attack when the
% real attack diffusion is high. Further, 
Building on top of
 previous related work \cite{Allodi-2014-TISSEC}, we show that patching policies based on our estimator show a comparable or better reduction in risk
than current best practices by requiring an essentially halved workload
in terms of patched vulnerabilities without losing ability of foiling attacks in the wild.

This paper is organised as follows: 
Section \ref{sec:datasets} introduces our datasets.
Section \ref{sec:trade-off} gives a first overview of exploits in Impact and Complexity. In Section \ref{sec:pa} we introduce
the \emph{Attack Potential} measure, and in Section \ref{sec:patch_policies} we evaluate our estimator against real attacks in the wild. Section \ref{sec:threats} discusses threats to validity; related work is
presented in Section \ref{sec:related}, and Section \ref{sec:conclusions} concludes the paper.

%%%%%%%%%%%%%%%%%%%%%%%%%%%%%%%%%%%%%%%%%%%%%%%%%%%%%%%%%%%%%%%%%%%%%%%%%%%%%
\section{Datasets} \label{sec:datasets}

Our analysis is based on three datasets:

%\begin{enumerate}
%\item 
(1) The National Vulnerability Database (\NVD) is usually considered the public
    universe of vulnerabilities
held by NIST.
Our \NVD\ samples
reports 49599 vulnerabilities (CVE identifiers).

%\item 
(2) To approximate a collection of vulnerabilities used in the wild we have used
Symantec's
AttackSignature\footnote{\url{http://www.symantec.com/security\_response/attacksignatures/}}
and
ThreatExplorer\footnote{\url{http://www.symantec.com/security\_response/threatexplorer/}}
public data (\SYM). It contains all entries identified as malware (local threats) or
remote attacks (network threats) by Symantec's commercial products. It reports
1277 vulnerabilities.

%\item 
(3) Symantec's \WINE\ data sharing program \cite{Dumitras-11-BADGERS} collects records of
attacks for attack signatures and CVEs reported in \SYM.
% We then exploit the mapping \{\emph{attack signature $\rightarrow$ CVE(s)}\} in
% \SYM\ to assign volumes of attacks to exploited vulnerabilities. 
Our \WINE\ sample has been collected in Summer 2012 and reports 2M attacks detected between August 2009 and June 2012 against $10^6$ systems\cite{Dumitras-11-BADGERS}, and exploiting 408 vulnerabilities.
%\end{enumerate}

\section{Explorative attack data analysis}
\label{sec:trade-off}

\begin{figure}[t]
\begin{center}
\includegraphics[width=0.45\textwidth]{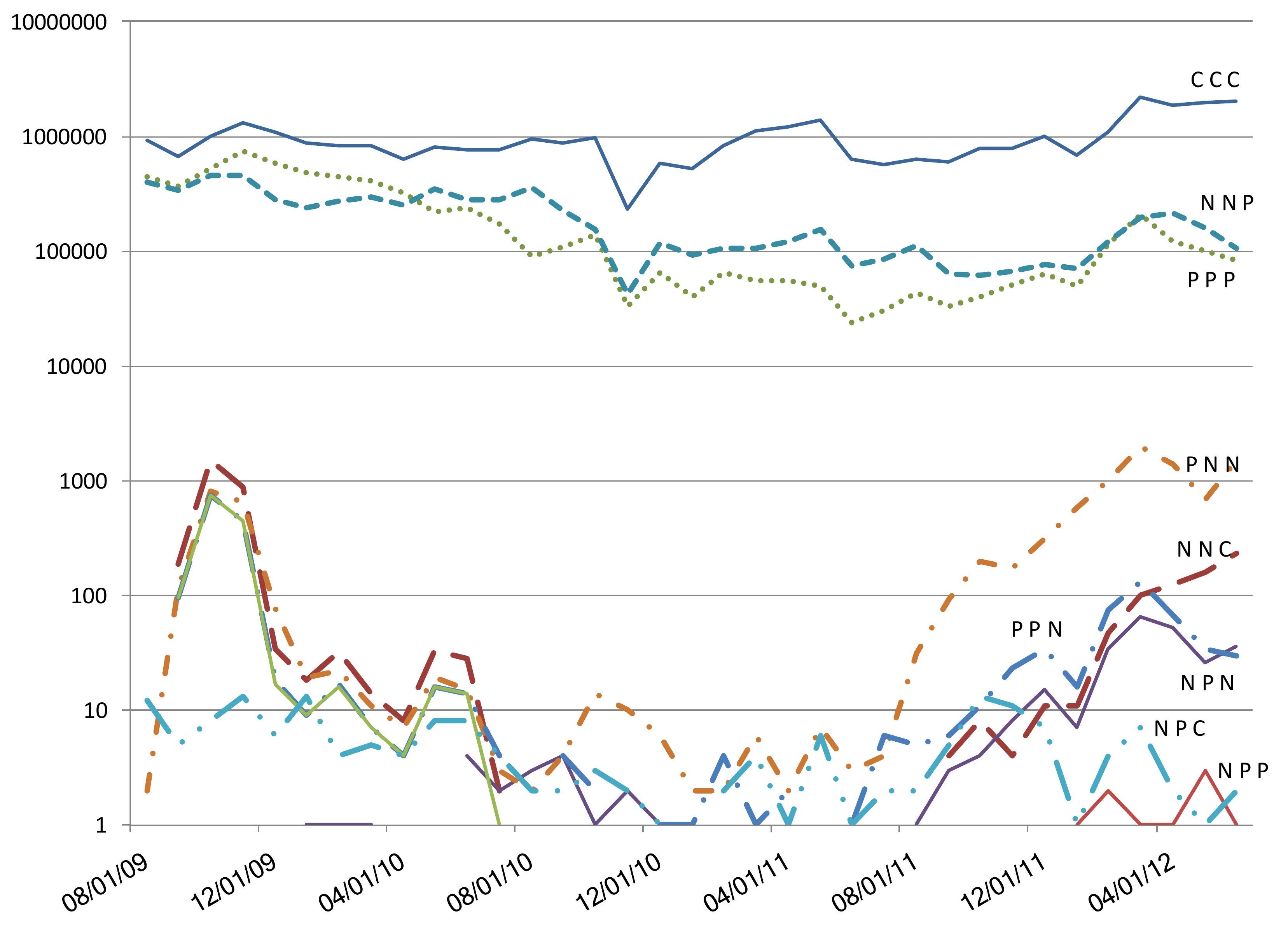}
\end{center}
\begin{scriptsize}
\end{scriptsize}
\begin{minipage}{0.95\columnwidth}
\footnotesize
\vspace{0.05in}
High \texttt{Impact} vulnerabilities are exploited several orders of magnitude more frequently than low \texttt{Impact} vulnerabilities.
\end{minipage}
\caption{Trends of attacks by impact type. Y-axis is in logarithmic scale}
\label{fig:wine-imp-trends}
\end{figure}

CVSS classifies vulnerability Impact over an assessment of the classic \emph{Confidentiality, Integrity, Availability} properties, expressed in terms of \emph{(C)omplete, (P)artial, (N)one} losses.
% can assume one of the values $C$ (complete), $P$ (partial) or $N$ (none). Therefore,
% a complete confidentiality only attack would be CNN, and one with complete
% integrity and partial availability would be NCP.
Figure \ref{fig:wine-imp-trends} reports the trend in volumes of attacks per CIA
impact type. Only CIA configurations for which there is an entry in \WINE\ are reported (the interested reader can refer to \cite{Allodi-2014-TISSEC} for a more detailed analysis of incidence of impact types on vulnerabilities and exploits).
The Y-axis is in logarithmic scale. It is immediately evident that two levels
of prevalence of
CIA impacts in attacks
can be identified:
vulnerabilities with C,I,A assessment $<$C,C,C$>$$<$P,P,P$>$,$<$N,N,P$>$ are up to 5 orders
 of magnitude \emph{more} exploited than most other impact types. 
 %The trend of
% attacks for the three most popular $<$C,I,A$>$ tuples 
% flattens on a logarithmic scale. 
$<$C,C,C$>$ and $<$P,P,P$>$ vulnerabilities
are among
the most targeted; this matches the observation that these impact types are among the most common overall in \NVD~\cite{Allodi-2014-TISSEC}. 
On the other hand, the high incidence of $<$N,N,P$>$ vulnerabilities in \WINE\ does not match a high presence of this impact type in \NVD~\cite{Allodi-2014-TISSEC}.
%  not only does this category
% identify a Low Impact (2.9 on a scale of 10), it is also not among
% the most common tuples for the score in \NVD~\cite{Allodi-2014-TISSEC}.
% Given this analysis, it seems that the CIA approach to vulnerability
% assessment can be greatly simplified without losing insight ability to classify
% impact types on attacks: the great majority of vulnerabilities targeted in
% cyber attacks
% allow for a \textsf{Complete Compromise}, \textsf{Partial Compromise} and/or
% \textsf{Crash} of the victim system.
% While it is immediately apparent that
% $<$C,C,C$>$ and $<$P,P,P$>$ vulnerabilities represent two
% of the highest-impact combinations for vulnerabilities,
% it is perhaps 
Further, it is useful to observe that $<$N,N,P$>$ vulnerabilities
are more commonly exploited in the wild than $<$N,N,C$>$
vulnerabilities, despite a lower overall impact (Partial Availability as opposed to Complete Availability impact).
% We further investigated this by looking at the characteristics
% of $<$N,N,P$>$ vulnerabilities. As reflected by
% the impact on Availability, these vulnerabilities
% may allow the attacker to perform
% denial-of-service attacks.
% But what differentiates `Partial Availability' Impact vulnerabilities from `Complete'
% ones, making them preferable to the latter?
% To investigate this, we consider the \emph{CVSS Exploitability} assessment
% of the vulnerabilities.
% Among the assessment that characterise \emph{Exploitability}, we
% consider \emph{\texttt{Access Complexity}}, an estimate of how difficult it is to reliably exploit
% the vulnerability \cite{Mell-2007-CMU}.
Some light can be shed on the apparent mismatch between volume of attacks and relative impact of affected vulnerabilities by considering the \texttt{Access Complexity}  levels reported in the \NVD. We find that 95\% of attacked $<$N,N,P$>$ vulnerabilities have a Low CVSS \texttt{Access Complexity}, and that more than 50\% of $<$N,N,C$>$ vulnerabilities are scored as High or Medium complexity.
This suggests that the combination of CVSS \texttt{Impact} and \texttt{Access Complexity} assessments may provide a useful first indicator of presence of exploit in the wild.

\paragraph{Trends of attacks in Impact and Complexity}
\label{sec:trends}

%We further investigate this relationship between impact scores
%and \emph{Exploitability} (sub)scores.
Following official guidelines~\cite{Mell-2007-CMU}, 
we categorise vulnerabilities by their impact and complexity over three levels for each metric: $HIGH, MEDIUM, LOW$. Vulnerability characteristics are
identified in short by the tuple $<$AC=X,I=Y$>$, with $x$ and $y$ the
assessments for the \texttt{Access Complexity} and \texttt{Impact} metrics respectively.
Table \ref{tab:cvss-expl} reports the relative fractions of vulnerabilities in \SYM\ for each combination. % of the Impact and Complexity measures.
\begin{table}[t]
\centering
\small
\caption{Relationship between \texttt{Access Complexity}, \texttt{Impact} and presence of exploit.}
\label{tab:cvss-expl}
\begin{minipage}{0.95\columnwidth}
\footnotesize
Fraction of vulnerabilities exploited in the wild reported in \SYM by \texttt{Impact} and \texttt{Access Complexity}. Low-complexity vulnerabilities are the most frequent ones. Medium-complexity vulnerabilities are only exploited when matched by a High \texttt{Impact}. High-complexity vulnerabilities are seldom exploited. Fractions in \SYM are clearly different from those in NVD, indicating that the selection process of vulnerability exploits is not random.
\vspace{0.05in}
\end{minipage}
\begin{tabular}{l l c c}
\toprule
 \texttt{Acc. Complexity} & \texttt{Impact} &\SYM & \NVD\\
\midrule
% \multirow{9}{*}{\begin{sideways}\texttt{Access Complexity}\end{sideways}}
   \multicolumn{1}{l}{\multirow{3}{*}{HIGH}} & HIGH &1.33\%&0.92\%\\
      \multicolumn{1}{l}{} &MEDIUM&1.88\%&1.89\%\\
     \multicolumn{1}{l}{} & LOW & 1.02\%&1.89\%\\[1ex]%\cline{3-5}
   \multicolumn{1}{l}{\multirow{3}{*}{MEDIUM}} & HIGH &32.50\%&7.65\%\\
     \multicolumn{1}{l}{} & MEDIUM & 3.60\%&7.69\%\\
     \multicolumn{1}{l}{} & LOW&2.43\%&14.83\%\\[1ex]%\cline{3-5}
   \multicolumn{1}{l}{\multirow{3}{*}{LOW}} & HIGH & 18.09\%&11.80\%\\
    \multicolumn{1}{l}{}  &MEDIUM & 22.55\%&30.43\%\\
    \multicolumn{1}{l}{}  &LOW&16.60\%&22.90\%\\ \bottomrule
\end{tabular}

\end{table}
It is evident that Low complexity and Medium complexity, High impact vulnerabilities are over-represented in \SYM\ with respect to \NVD. For example, $<$AC=M, I=H$>$ vulnerabilities constitute 32.5\% of vulnerabilites in \SYM, whereas they represent only 7.65\% of vulnerabilities in \NVD. Similarly, $<$AC=M, I=\{M,L\}$>$ vulnerabilities are under-represented in \SYM\ with respect to occurrences in \NVD.

\begin{figure}[t]
\begin{center}
\includegraphics[width=0.45\textwidth]{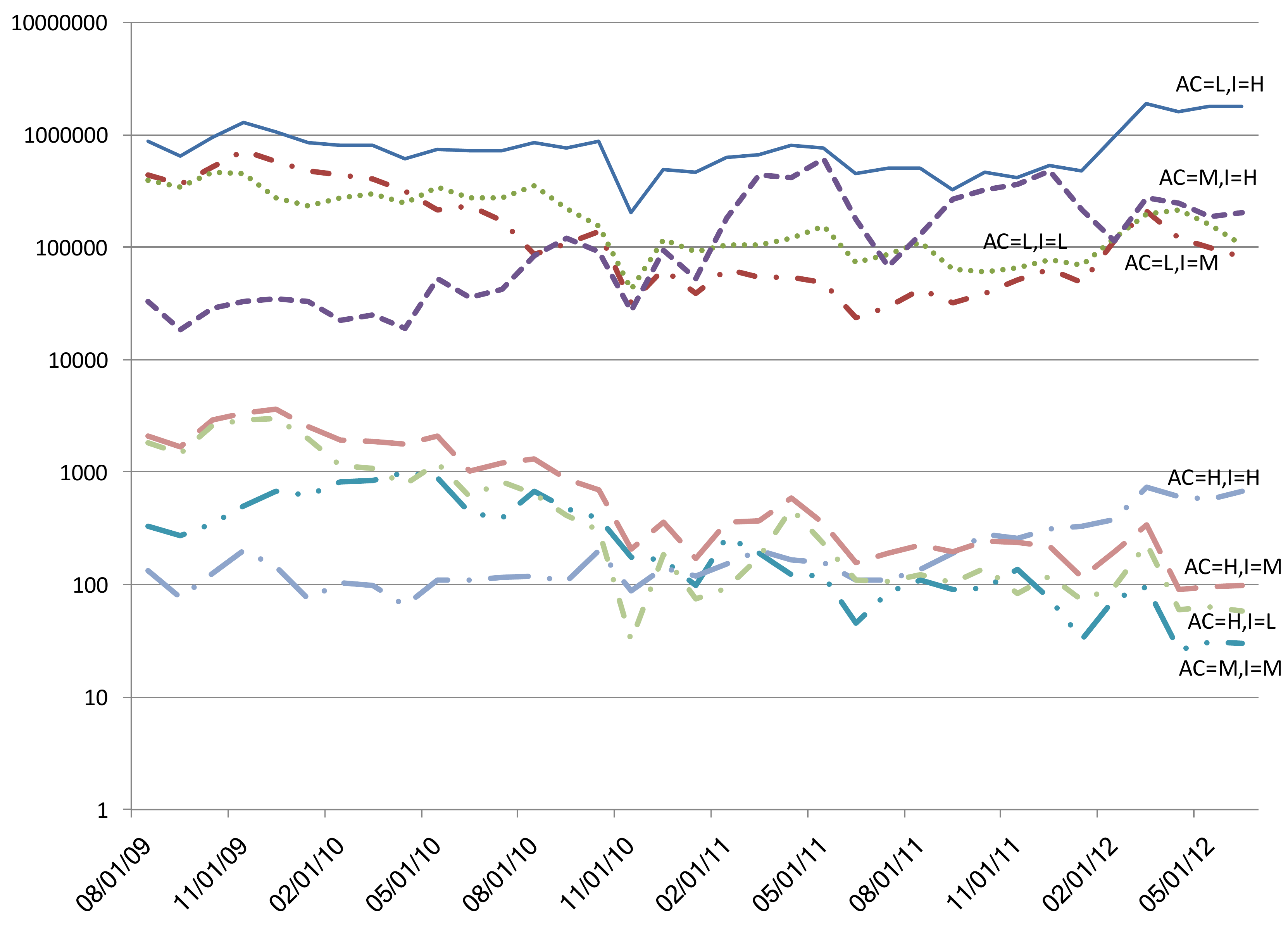}
\end{center}
\begin{minipage}{0.95\columnwidth}
\footnotesize
\vspace{0.05in}
Vulnerabilities with a high impact-complexity trade-off are exploited orders of magnitude more often than other vulnerabilities. Medium \texttt{Acc. Complexity} vulnerabilities drive attacks only when matched by a high \texttt{Impact}.
\end{minipage}
\caption{Attacker trade-offs in Impact vs Complexity types. Y-axis is in logarithmic scale.}
\label{fig:wine-imp-vs-compl}
\end{figure}

In 
Figure \ref{fig:wine-imp-vs-compl} we report attack volumes
in time aggregated
by \{\emph{complexity, impact}\} types. 
The Y-axis is in logarithmic scale.
In the plot the $<$AC=M, I=L$>$ tuple is not reported
as many data points in the time series are zeros.
The great majority of exploited vulnerabilities are Low \texttt{Access Complexity}
ones; among these, the most attacked are High \texttt{Impact} vulnerabilities.
Medium \texttt{Access Complexity} vulnerabilities are massively exploited only if their impact on the victim systems is High. 
The resulting picture 
is therefore rather simple: the majority of exploits are for easy vulnerabilities to exploit,
regardless of their Impact type. Medium complexity vulnerabilities
are less targeted (by two orders of magnitude), and only if the  
exploitation impact is High.
Essentially we can see that Low Complexity and High Impact vulnerabilities remain at large the favorite vector for attacks while the remaining Low Complexity or Medium Complexity but High Impact are still very popular albeit by varying degrees (oscillating in 1 order of magnitude below the top category). The other vulnerability types remain below at several orders of magnitude. This suggests that the combination of \texttt{Access Complexity} and \texttt{Impact} may help identifying useful measures for incidence of attacks. %The identified trends are substantially stable in time.

% We identify two levels of \emph{exploitation profitability}: \emph{High
% profitability} includes $<$C=L,I=*$$>$ and $<$C=M,I={H}$$>$ 
% vulnerabilities, i.e. vulnerabilities
% that have a more favourable pay-off for the attacker.
% \emph{Low profitability} vulnerabilities are in the groups 
% $<$C=M,I=\{L,M\}$$>$ and
% $<$C=\{H\},I=*$>$, for which
% either too high complexity or too low final impact
% lower the attacker's pay-off.
% % The gap between the two \emph{profitability} levels increases with time, highlighting the
% % uprising difference in importance of these two categories of vulnerabilities.
% We find \emph{High profitability} vulnerabilities to be
% attacked up to \emph{four orders of magnitude more} than \emph{Low profitability} ones.

% To understand how many vulnerabilities are responsible for the two levels of
% `profitability', we report in 

% The dominance of \emph{high profitability} vulnerabilities
% is evident in the LOW and MEDIUM \texttt{Access Complexity} types. The 
% trade-off is particularly evident in the latter, where only HIGH impact 
% vulnerabilities result to be of interest for attackers.

\section{Potential of attack}
\label{sec:pa}
\begin{figure}[t]
\centering
\includegraphics[width=0.4\textwidth]{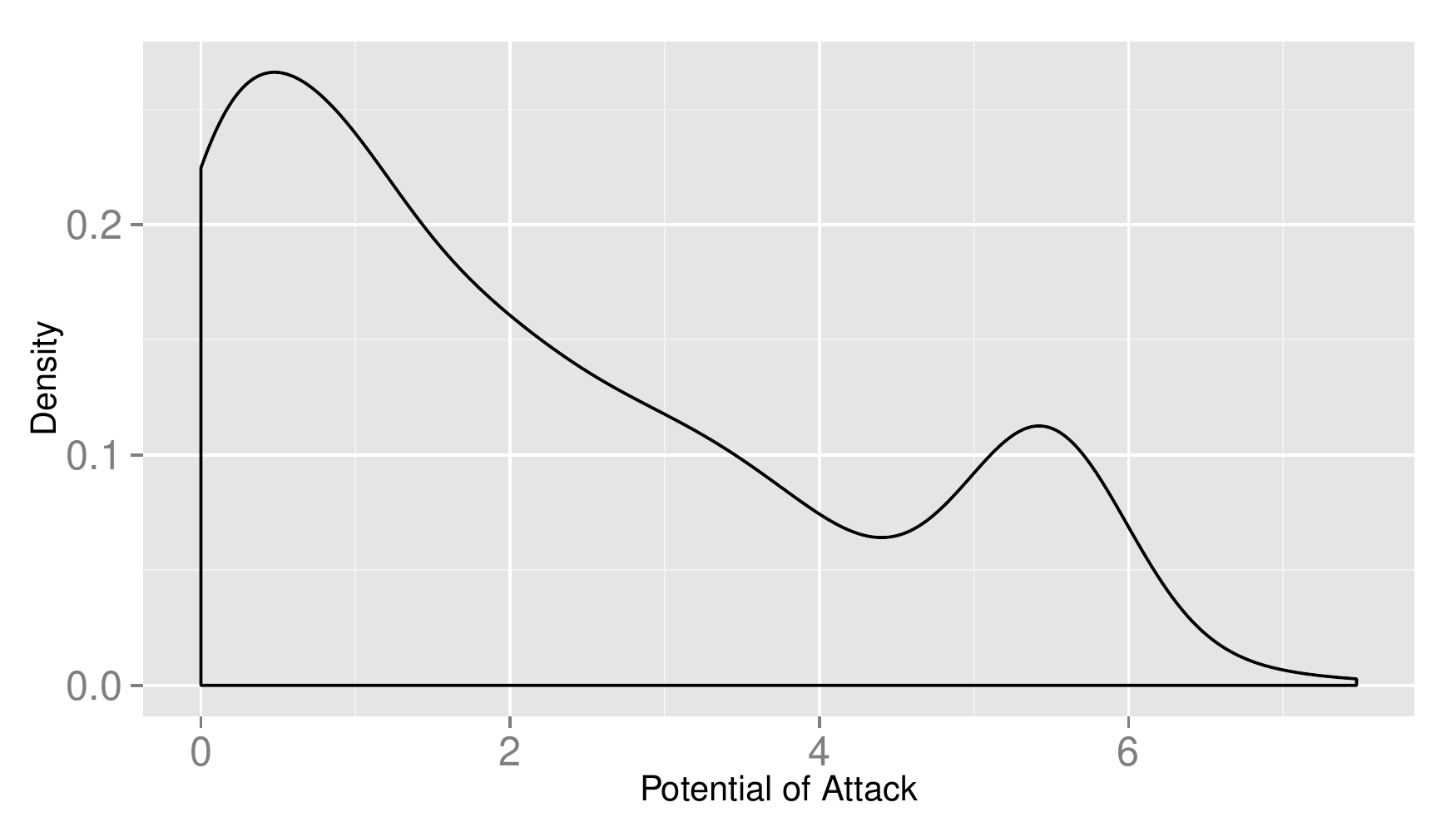}
\caption{Probability density of $pA$ measures in \WINE.}
\label{fig:pa-density}
\end{figure}

To more precisely describe the trends outlined in Section \ref{sec:trends}
we % introduce the notion of `potential of attack', $pA$ in short.
define $pA$ as an empirical measure of the \emph{potential of attack}
of a vulnerability over a set of vulnerable machines.\footnote{In chemistry, the $pH$ of a solution is a function of the (molar) concentration
 of hydrogen ions. Similarly, in our definition of $pA$ we consider the presence of attacks in the wild recorded over the \WINE\ sample.}
 % To be precise, it is an 
 % \emph{empirical measure} of
 % the capacity of the hydrogen ions
 % to be involved in chemical reactions (and therefore determine the degree of 
 % acidity of a solution). Because the
 % concentration of these ions is typically low, $pH$
 % is calculated as the logarithm of the inverse of $[H^+]$.
 % More formally, $pH=log_{10}\frac{1}{[H^+]}$.}.
%To compute $pA$ we rely on our \WINE\ sample. %, that comprises attack records for $10^7$ machines.
% a set of $10^7$ machines\footnote{In this way with a $pA=7$ the number of attempted attacks is globally equal to the number of machines. Obviously some machines may be attacked more than once and some machines may not be attacked at all.}. 
Our $pA$ measure is specified as
\begin{equation}
pA = log_{10}(A_v)
\end{equation}
where $A_v$ is the number of attacks observed in the wild for the vulnerability $v$. 
A $pA\ \text{of}\ 6$ corresponds to one million attacks in the wild in our sample (i.e. one per systems on average). A $pA=2$ indicates 100 recorded attacks.\footnote{Obviously, over a set of $10^6$ machines some may be attacked more than once and some may not be attacked at all.}
% Notice that we cannot use the attack potential as a decision factor because when its value is known it would be too late. In medical terms this would be equivalent to consider as a risk factor for getting cancer the very fact that we\ldots already have it! The risk factors, such as the presence on black markets or the CVSS score, can instead be estimators for the attack potential.
Figure~\ref{fig:pa-density}
reports the probability density distribution of $pA$ for vulnerabilities reported in
 \WINE. The x axis reports
the $pA$ values and the y axis the incidence of each $pA$. $pA$ ranges
in $[0..7.5]$. Its distribution spikes at $pA\approx1$
and $pA\approx6$. 50\% of vulnerabilities are below a $pA$ of 1.6,
and 75\% below 3.4. This means that 50\% and 75\%
of vulnerabilities receive respectively up to about 50 
($10^{1.6}$) and 2500 ($10^{3.4}$) attacks in the wild over the observation period.

 \paragraph{Estimation of vulnerability attack potential}
 \label{sec:estimator}

% Results from Figure \ref{fig:wine-imp-vs-compl}
% and 
% Table \ref{tab:cvss-expl} show that there is a clear correlation between  Impact and Attack Complexity levels of a vulnerability and the presence of an attack.
We exploit the impact-complexity effect described in Section~\ref{sec:trends} to estimate the volume of attacks that a 
vulnerability can potentially receive if an attack for it exists in the wild.
%This can be particularly useful in practice when the CVSS assessment is used 
%to prioritize patching work and/or assess vulnerability risk (e.g. notably: 
%NIST SCAP protocol \cite{Scarfone-2010-SCAP}, PCI DSS \cite{PCI-DSS-DOC}).
% From the analysis in Section \ref{sec:trends} we derive an estimator
% for $pA$ based on the Impact and \texttt{Access Complexity} assessments. 
Given the high incidence of unexploited vulnerabilities in the wild~\cite{Nayak-2014-RAID}, a desirable property for our estimator is to maintain high true negative rates (something that the bare \texttt{CVSS-score} unfortunately does not do~\cite{Allodi-2014-TISSEC}), whereas false positives can be ruled out by more fine-grained assessments later in a triage process~\cite{chakradeo2013mast}.
To build our estimator, we first
assign to each \texttt{Impact} and \texttt{Access Complexity} value an ordinal value
derived directly from the original CVSS v2 specification \cite{Mell-2007-CMU}.
Table \ref{tab:formulaScores}
\begin{table}
\centering
\small
\caption{Scores by \texttt{Impact} and \texttt{Access Complexity} type. Values
derived from the original CVSS v2 formula \cite{Mell-2007-CMU}.}
\label{tab:formulaScores}
\begin{tabular}{lr}
\toprule
\texttt{Impact} & Score \\\midrule
% CCC &10\\
% CCP &9\\
% CCN &9\\
% PPP &6\\
% PPN &6\\
% NPN &3\\\hline
H & 10\\
M & 6 \\
L & 3 \\\bottomrule
\end{tabular}
\begin{tabular}{lr}
\toprule
\texttt{Acc. Complexity} & Score\\\midrule
L &10\\
M &7\\
H &2\\\bottomrule
\end{tabular}
\end{table}
reports the assigned scores. Leveraging the logarithmic
relation between attacks in the wild 
and CVSS measures, we define the estimated
attack potential $E[pA]$ as:

\begin{equation}
E[pA] = log_{10}(Impact) \times (Complexity)
\label{eq:estimator}
\end{equation}

The estimator in Eq. \ref{eq:estimator} will return values of
$E[pA] \in [0,10]$. Because our \WINE\ set is collected
over $10^6$ machines~\cite{Dumitras-11-BADGERS}, we define the following discrete levels:

\begin{itemize}
\item $5<E[pA]$: HIGH
\item $3<E[pA]\leq5$: MEDIUM
\item $0\leq E[pA]\leq3$: LOW
\end{itemize}

% \subsection{Example of estimation}
% \label{sec:example}

% Table \ref{tab:cveDescriptions} reports an estimation of the
% % \begin{table*}
% % \centering
% % \footnotesize
% % \caption{Example vulnerabilities sampled from WINE}
% % \label{tab:cveDescriptions}
% % \begin{tabular}{llcccc}
% % %&&\multicolumn{4}{c}{CVSS v2 Assessment}\\
% % \multicolumn{1}{l}{cve}&software&\texttt{Access Complexity}&Confidentiality&Integrity&Availability\\\hline
% % \multicolumn{1}{l}{CVE-1999-0749}\\
% % \multicolumn{1}{l}{CVE-1999-1519}\\
% % \multicolumn{1}{l}{CVE-2003-0109}\\
% % \multicolumn{1}{l}{CVE-2005-4459}\\
% % \multicolumn{1}{l}{CVE-2007-0015}\\
% % \multicolumn{1}{l}{CVE-2007-0038}\\
% % \multicolumn{1}{l}{CVE-2008-0639}\\
% % \multicolumn{1}{l}{CVE-2008-0726}\\
% % \multicolumn{1}{l}{CVE-2008-5359}\\
% % \multicolumn{1}{l}{CVE-2009-0075}\\
% % \multicolumn{1}{l}{CVE-2009-3882}\\
% % \multicolumn{1}{l}{CVE-2009-3886}\\
% % \multicolumn{1}{l}{CVE-2010-0806}\\
% % \multicolumn{1}{l}{CVE-2010-2265}\\
% % \multicolumn{1}{l}{CVE-2010-3974}\\\hline 
% % \end{tabular}
% % \end{table*}
% a sample of twelve vulnerabilities chosen at random from \WINE.
% % that we will be instrumental to illustrate our estimator.
% The column software reports the name of the first affected software reported by the \NVD. 
% The full verbose description of the vulnerabilities is reported in Table
% \ref{tab:appendix}, in the Appendix.
As an application example of our measure, Table \ref{tab:pAEstimates}
\begin{table*}[t]
\centering
\small
\caption{$E[pA]$ estimates for vulnerabilities in the WINE sample.}
\label{tab:pAEstimates}
\begin{minipage}{0.9\textwidth}
\footnotesize
Calculation of $E[pA]$ values for ten vulnerabilities in WINE. The estimator only need in input \texttt{Access Complexity} and \texttt{Impact} values from the \texttt{CVSS-vector}.
\vspace{0.05in}
\end{minipage}
\begin{tabular}{llcc |r| rl| rl|c}
\toprule
\multicolumn{5}{c}{Vulnerability characteristics}&\multicolumn{2}{c}{$pA$}&
\multicolumn{2}{c}{$E[pA]$}&\multicolumn{1}{c}{}\\[0.03in]
CVE&Sw&AC&Impact&\#attacks&$pA$&Discrete $pa$&$E[pA]$&Discrete $E[pA]$&$E[pA]\approx pA$\\\midrule
CVE-2010-0806&internet\_explorer&M&H&126765&5.1&HIGH&7&HIGH&\tick\\
CVE-2009-3886&jre&L&M&427315&5.6&HIGH&8.4&HIGH&\tick\\
%CVE-2009-3882&jdk&L&M&427&5.6&HIGH&8.4&HIGH&\tick\\
CVE-2009-0075&internet\_explorer&M&H&2803&3.4&MEDIUM&7&HIGH&$E[pA]>pA$\\
CVE-2008-5359&jdk&M&H&377324&5.6&HIGH&7&HIGH&\tick\\
CVE-2008-0726&acrobat&M&H&246049&5.3&HIGH&7&HIGH&\tick\\
%CVE-2008-0639&client&L&H&8&0.9&LOW&10&HIGH&$E[pA]>pA$\\
CVE-2007-0038&windows\_2000&M&H&2291&3.3&MEDIUM&7&HIGH&$E[pA]>pA$\\
CVE-2007-0015&quicktime&M&M&147857&5.1&HIGH&5.9&HIGH&\tick\\
CVE-2005-4459&ace&L&H&4&0.6&LOW&10&HIGH&$E[pA]>pA$\\
CVE-2003-0109&windows\_2000&L&M&203530&5.3&HIGH&8.4&HIGH&\tick\\
CVE-1999-0749&windows\_95&H&L&20&1.3&LOW&0.9&LOW&\tick\\\bottomrule
% CVE&SW&AC&C&I&A&\#attacks&pA&Discrete pA&$E[pA]$&Discrete $E[pA]$&$E[pA]\approx pA$\\\hline
% CVE-1999-0749&windows\_95&H&N&P&N&21&1.3&LOW&1&LOW&\tick\\%\cline{2-7}
% CVE-1999-1519&g6\_ftp\_server&L&N&N&P&7&0.8&LOW&4.8&MEDIUM&\tick\\%\cline{2-7}
% CVE-2003-0109&windows\_2000&L&P&P&P&203530&5.3&HIGH&7.8&HIGH&\tick\\%\cline{2-7}
% CVE-2005-4459&ace&L&C&C&C&5&0.7&LOW&10&HIGH&Too high\\%\cline{2-7}
% CVE-2007-0015&quicktime&M&P&P&P&147857&5.2&HIGH&5.4&HIGH&\tick\\%\cline{2-7}
% CVE-2007-0038&windows\_2000&M&C&C&C&2291&3.4&MEDIUM&7&HIGH&\tick\\%\cline{2-7}
% CVE-2008-0639&client&L&C&C&C&9&1&LOW&10&HIGH&Too high\\%\cline{2-7}
% CVE-2008-0726&acrobat&M&C&C&C&246049&5.4&HIGH&7&HIGH&\tick\\%\cline{2-7}
% CVE-2008-5359&jdk&M&C&C&C&377324&5.6&HIGH&7&HIGH&\tick\\%\cline{2-7}
% CVE-2009-0075&internet\_explorer&M&C&C&C&2803&3.4&MEDIUM&7&HIGH&\tick\\%\cline{2-7}
% CVE-2009-3882&jdk&L&P&P&P&427315&5.6&HIGH&7.8&HIGH&\tick\\%\cline{2-7}
% CVE-2009-3886&jre&L&P&P&P&427315&5.6&HIGH&7.8&HIGH&\tick\\%\cline{2-7}
% CVE-2010-0806&ie&M&C&C&C&126765&5.1&HIGH&7&HIGH&\tick\\%\cline{2-7}
% CVE-2010-2265&windows\_2003\_server&M&N&P&N&9432&4&MEDIUM&3.3&MEDIUM&\tick\\%\cline{2-7}
% CVE-2010-3974&windows\_2003\_server&H&C&C&C&2&0.3&LOW&2&MEDIUM&\tick\\\hline
\end{tabular}
\end{table*}
reports the $pA$ and $E[pA]$ estimates for ten example vulnerabilities randomly sampled from \WINE.
Table~\ref{tab:epavspa}
\begin{table}
\centering
\small
\caption{$E[pA]$ and $pA$ discrete estimations.}
\label{tab:epavspa}
\begin{minipage}{0.95\columnwidth}
\footnotesize
Our $pA$ estimator conservatively estimates real volumes of attacks in the wild, and seldom under-estimates real $pA$ values. A low $E[pA]$ correctly matches a low-$pA$ vulnerability 67\% of the time, whereas medium and high $E[pA]$ levels seem to consistently over-estimate low-$pA$ vulnerabilities. 
\vspace{0.05in}
\end{minipage}
\begin{tabular}{l l r r r | r}
\toprule
&& \multicolumn{4}{c}{ $E[pA]$}\\[0.05in]
&&HIGH&MEDIUM&LOW&\textbf{Sum}\\\midrule
\multirow{3}{*}{\rotatebox[origin=c]{90}{ $pA$}}&
% HIGH&68&1&17&\textbf{86}\\
% &MEDIUM&62&2&7&\textbf{71}\\
% &LOW&204&4&43&\textbf{251}\\
% &\textbf{Sum}&\textbf{334}&\textbf{7}&\textbf{62}&\textbf{408}\\
%&HIGH&MEDIUM&LOW&Sum\\
HIGH&70&13&3&\textbf{86}\\
&MEDIUM&62&2&7&\textbf{71}\\
&LOW&197&34&20&\textbf{251}\\\midrule
&\textbf{Sum}&\textbf{329}&\textbf{49}&\textbf{30}&\textbf{408}\\
\bottomrule
\end{tabular}
\end{table}
reports overall results for the full \WINE\ dataset. The estimator performs generally well, with only 6\% of the assessments being \emph{under}-estimations of real $pA$ (i.e. false negatives). Importantly, Table~\ref{tab:epavspa} shows that our estimator is a \emph{conservative} one, in that when it does not match the correct $pA$ category, it \emph{over}-estimates it. It therefore does not lead to ignoring vulnerabilities that should be treated. Among the false-negatives, 87\% of the estimation error is limited to one level only on the discrete scale. Note that because all vulnerabilities considered in Table~\ref{tab:epavspa} have at least an exploit in the wild (as they are reported in \WINE), the reported indicators do not represent real-world performance of the estimator. We give full a consideration of this in the next section.
%  is reflected in an \emph{over}-estimation of the real $pA$.
% The estimator for $pA$ works generally well. The
% estimated value is typically close to the real one, with legligible most `misses'
% being `overshoots' that are however missing the estimate by only one
% level on the discrete scale. Importantly, the estimator does not
% underestimates the criticality of a vulnerability, i.e. 
% the estimator does not miss vulnerabilities that represent
% high risk.

\section{An application to illustrative patching policies}\label{sec:patch_policies}

% \section{Discussion}
% \label{sec:discussion}

% In this paper we show that part of the attacker
% decision process in choosing which vulnerabilities to attack
% \emph{en-masse} is ruled by a simple trade-off mechanism
% whereby vulnerabilities with a high impact and low exploitation complexity
% are preferred by the attacker.

% This is reflected in both the number of exploited vulnerabilities
% and in the volume of attacks
% received for each \emph{Impact,Complexity} tuple.
% We leverage this effect to build an estimator that, given the
% CVSS assessment of the vulnerability, can discern between vulnerabilities
% with a high potential of attack and vulnerabilities with a low potential of
% attack. This can be particularly useful in practice to prioritise 
% patching work, as vulnerabilities with a high attack potential
% represent higher risk and therefore should be fixed with higher priority.
% The estimator presented in this work is straightforward to implement
% and can be used in foresight without pre-existing data on the
% volume of attacks in the wild for that vulnerability.

% \subsection{Illustrative Patching Policies}

To illustrate the practical application of our estimator to vulnerability management practices, we define a patching policy as a process that, given
in input vulnerability data, outputs a \emph{Patch/NotPatch} decision. 
A patching policy defines a \emph{threshold} above which the
\emph{Patch} decision is triggered.
We define the following policies:
% We combine each of these risk factors with all \emph{Base policies} and 
% evaluate them separately.
\begin{itemize}
\item $All\ vulns$: no risk factor is identified; under this
policy every vulnerability is patched.
    \item $CVSS\geq 4$: patches all vulnerabilities
    to which is assigned a CVSS score equal or higher than 4. This policy
    corresponds to the PCI DSS recommendation for management  of credit card holders data \cite{PCI-DSS-DOC}.
% \item $CVSS\geq 9$: patches vulnerabilities 
%  only if the
% CVSS score is greater than 9.
\item $COMPL=L$: patches `easy to exploit
vulnerabilities' with a CVSS Complexity assessment = $L$.
\item $COMPL\leq M$: patches only `low hanging fruits' vulnerabilities
with a CVSS Complexity assessment either $L$ or $M$. High complexity 
vulnerabilities are ignored.
\item $E[pA]=H$: accounts for vulnerabilities with a high estimated $pA$ ($E[pA]>5$).
\end{itemize}

% Table~\ref{tab:workload} reports vulnerabilities marked to patch by the identified policies.
% % \REMARK{Fabio}{PUT HERE A TABLE WITH THE COST ONLY.}
% \begin{table}[t]
% \centering
% %\footnotesize
% \caption{No. of vulnerabilities to fix by policy.}\label{tab:workload}
% \begin{tabular}{lr}
% Policy &  \#V\\\hline
% \emph{All vulns} & 14380\\
% $CVSS\geq4$& 13715\\
% %$CVSS\geq4 + PoC$& 3004\\
% %$CVSS\geq4 + Black$& 58\\\hline
% %$CVSS\geq9$& 3081\\
% %$CVSS\geq9 + PoC$& 550\\
% %$CVSS\geq9 + Black$& 48\\\hline
% $COMP=L$& 7552\\
% %$COMP=L + PoC$& 1964\\
% %$COMP=L + Black$& 21\\\hline
% $COMP\leq M$& 13952\\
% %$COMP\leq M + PoC$& 2975\\
% %$COMP\leq M + Black$& 56\\\hline
% $E[pA]=H$& 8308\\\hline
% %$COMP\leq M \& Imp\geq M + Exist$& 2408\\
% %$COMP\leq M \& Imp\geq M + Black$& 53\\\hline\hline
% % \emph{Exist} & 3539\\
% % \emph{PoC} & 400 \\
% % \emph{Black} & 58 \\\hline
% % $CVSS\geq4 + Exist$& 3378\\
% % $CVSS\geq4 + PoC$& 395\\
% % $CVSS\geq4 + Black$& 58\\\hline
% % $CVSS\geq9 + Exist$& 1652\\
% % $CVSS\geq9 + PoC$& 238\\
% % $CVSS\geq9 + Black$& 48\\\hline
% % $COMP=L + Exist$& 1469\\
% % $COMP=L + PoC$& 175\\
% % $COMP=L + Black$& 21\\\hline
% % $COMP\leq M + Exist$& 3392\\
% % $COMP\leq M + PoC$& 391\\
% % $COMP\leq M + Black$& 56\\\hline
% % $COMP\leq M \& Imp\geq M + Exist$& 2280\\
% % $COMP\leq M \& Imp\geq M + Exist$& 314\\
% % $COMP\leq M \& Imp\geq M + Black$& 53\\\hline\hline
% \end{tabular}
% \end{table}

\subsection{Calculation of Risk Reduction}\label{sec:risk_reduction}

To evaluate the efficacy of the different patching
policies, we compute the frequency with which
each policy identifies a vulnerability in \SYM.
% Table \ref{tab:categories} 
% describes the software categories we identify
% \begin{table}[t]
% \centering 
% \footnotesize
% \caption{Categories used to classify vulnerabilities.} \label{tab:categories}
% \begin{tabular}{ l  l  l }
% Category & Type of software & Examples\\
% \hline
% \IE & IE browser & Internet Explorer\\[1ex]
% \PLU & Browser plugins & Acrobat reader\\
% && Flash Player\\[1ex]
% \COM & Productivity software & Microsoft Office\\
% && Eudora\\[1ex]
% \WIN & MS Windows releases & Windows XP\\
% && Vista\\\hline
% % \DEV & Software for developers & Visual C++ \\\hline
% % \BUS & Software used mainly in business environment & Lotus Notes, Dreamweaver\\\hline
% % \SER & Server side software & Apache, Ftp daemons\\\hline
% % \OTH & Other O.S. than MS Windows & Solaris, OpenBSD\\\hline
% % \NUL & Uncategorised software & Banner generator, Spip\\
% % \hline\hline
% \end{tabular}
% \end{table}
 %alongside a few explanatory examples of classified software. 
To evaluate each risk policy we randomly sample a
vulnerability from \NVD\ with the same characteristics
in terms of software and year of disclosure as in \SYM~\cite{Allodi-2014-TISSEC,Christey-2013-BHUSA}. We then evaluate the count of sampled vulnerabilities
that the policy marks as `high risk' (i.e. above
the threshold identified by the policy), and compare that
to the vulnerability's actual presence in \SYM.

The output of our experiment
is represented in Table \ref{tab:experiment-output}.
\begin{table}[t]
\centering
\small
\caption{Output format of our experiment.} \label{tab:experiment-output}
\begin{tabular}{l c c}
\toprule
 & $v\in\SYM$ & $v\not\in \SYM$\\\midrule
Above Threshold & a & b\\
Below Threshold & c & d\\
\bottomrule
\end{tabular}
\end{table}
% is a sample output of an experiment
% run.
% In order to measure the \emph{risk}  use the same data of
% Table~\ref{tab:examplesensispeci} but by row instead of by column. 
The first
row identifies the vulnerabilities that need be treated
 according to the decision
variables. 
The risk entailed by selected vulnerabilities is computed
on the first row, and is the ratio $R_{treated}=a/(a+b)$. The bottom row
identifies the vulnerabilities that are not selected for treatment
(below the identified threshold). The risk 
associated with the untreated vulnerabilities is the ratio 
$R_{untreated}=c/(c+d)$. The difference
between the two is defined in the literature as \emph{risk reduction (RR)} 
\cite{Evans-1986-ACC}. For a fixed number of vulnerabilities to patch, policies with a higher risk reduction identify a greater fraction of exploited vulnerabilities than policies with a lower risk reduction.

Formally, let $Attacked$ be the set of vulnerabilities for which attacks in the wild have been reported and $Selected$ the set of vulnerabilities above a policy's
threshold. The risk of treated (untreated) vulnerabilities for a patching policy $R_{treated}$ ($R_{untreated}$) and the \emph{risk reduction ($RR$)} of a policy are defined as:
\begin{eqnarray}
R_{treated} & = & \frac{|Attacked\cap Selected|}{|Selected|} \\
R_{untreated} & = & \frac{|Attacked\cap \neg Selected|}{|\neg Selected|} \\
RR & = & R_{treated} - R_{untreated}
\end{eqnarray}

% The risk reduction calculated in this way by the case-control study is an approximation of the actual risk reduction that we could calculate with a traditional prospective experiment. We should have indeed calculated the risk of the attacks that would have been suffered if the policy had not been 
% applied. Since this is a retrospective study we cannot obviously do that because some policy have been indeed applied in the past. For the statistically inclined \cite{evans1986double} provides an explanation of why this is a satisfactory approximation.
To implement
the procedure we perform a bootstrapped  
case-control study as
described in \cite{Allodi-2014-TISSEC}. Because
different software types may lead to different
attack frequencies, we identify four software
categories in \SYM\ to control for in our sample \cite{Allodi-2013-IWCC,Christey-2013-BHUSA}: \IE, \PLU, \COM\ and \WIN. The classification is performed by manually assigning software names to a category and then using regular expressions to match each CVE to the respective category.

\subsection{Risk reduction in the wild}

With this formulation we can compare the policies identified in Section~\ref{sec:patch_policies}.
Table~\ref{tab:full_results}
\begin{table}[t]
\centering
\small
\caption{Risk reduction and vulnerabilities to be considered per policy.}\label{tab:full_results}
\begin{minipage}{0.95\columnwidth}
\footnotesize
High risk reduction patching policies address a higher rate of exploited vulnerabilities by decreasing the rate of false positives. A policy based on our $E[pA]$ measure outperforms standard patching policies based on the sole \texttt{CVSS-score} or atomic values of the \texttt{CVSS-vector}.
\vspace{0.05in}
\end{minipage}
\begin{tabular}{p{3.5cm}rrr}
\toprule
Policy&\#V&RR&C.I.\\\midrule
$All\ vulns$ &14.380&-&-\\
$CVSS \geq 4$&13715&23.2\%& 21.2\% - 24.5\%\\
%$CVSS \geq 9$&3081&8.9\%& 8.5\% - 9.3\%\\
$Comp=L$&7552&13.0\%&11.9\% - 14.1\%\\
$Comp \leq M$&13952&5.2\%&2.6\% - 6.3\%\\
$E[pA]=H$&8308&26.8\%&24.5\% - 27.4\%\\\bottomrule
% $PoC$ &35.9\%&3.030&34.6-37.4\%\\
% $PoC+ CVSS \geq 4$&35.9\%&3004&34.4-37.5\%\\
% $PoC+ CVSS \geq 9$&42.0\%&550&39.9-43.1\%\\

% $PoC+ Comp=L$&31.2\%&1,964&29.9-32.6\%\\
% $PoC+ Comp \leq M$&35.8\%&2975&34.6-36.8\%\\
% $PoC+ Comp \leq M \& Imp \geq M$&37.7\%&2408&35.9-38.6\%\\\hline

% $Black$ &46.5\%&58&45.4-47.4\%\\
% $Black+ CVSS \geq 4$&45.4\%&58&44.4-45.8\%\\
% $Black+ CVSS \geq 9$&43.7\%&48&41.4-45.4\%\\
% $Black+ Comp=L$&37.0\%&21&32.2-38.2\%\\
% $Black+ Comp \leq M$&45.6\%&56&44.3-46.4\%\\
% $Black+ Comp \leq M \& Imp \geq M$&42.8\%&53&41.9-43.5\%\\
\end{tabular}
\end{table}
reports the results of the analysis. The first row reports the results
for the $All\ vulns$ policy. With reference to
our \NVD\ dataset sample, 
this would require analyzing more than 14 thousand vulnerabilities\footnote{The implementation of these policies
may require different levels of effort or costs. For example, the same vulnerability could be present in hundreds of machines or could reside in a server for which a 1 hour downtime is already too much. This information is company dependent and we do not consider it here. 
% We discuss in Section~\ref{sec:evaluation} how the whole framework can be lifted to consider the actual number of software instances with the vulnerability (and adjust the risk notion accordingly). 
Rather, we consider a simpler proxy information that is the number of vulnerabilities that are marked for patching by each policy.}. 
Risk Reduction can not be computed for this policy as no vulnerability remains unselected for patching. 
Among all policies, $Comp=L$ and $Comp \leq M$ achieve the lowest
risk reductions. This example is however useful in better illustrating the mechanism implemented by the RR metric. Whereas
$Comp=L$ is a \emph{sub}set of $Comp \leq M$, its risk-reduction is significantly higher; this may result counter-intuitive as the latter contains all vulnerabilities included in the former. However, note that by including additional vulnerabilities that are not attacked, $Comp \leq M$ decreases $R_{treated}$, thus resulting in a lower overall $RR$. A high $RR$ results therefore from policies that well balance the risk of selected vulnerabilities with the `residual' risk that characterizes unselected vulnerabilities.
The largest risk reduction is achieved by the policy based on
the impact-complexity trade-off ($RR=26.8\%$). The second
largest RR is achieved by a policy based on PCI-DSS. The former requires to analyze 8.3 thousand vulnerabilities and the latter almost 14 thousand. 
From this perspective the policy $E[pA]=H$ seems to have the best trade-off: lowest number of vulnerabilities and best risk reduction.

% We further check the impact of the software categories on the risk reduction
% measure.
Table~\ref{tab:results} reports the risk reduction for the aggregate case
and the experiment for the controls. Comparisons are by row.
\begin{table*}[t]
\caption{Risk reduction and workload, expressed in terms
of no. vulnerabilities patched, for all controls.}\label{tab:results}
\centering
\small

\begin{minipage}{0.9\textwidth}
\footnotesize
Risk reduction and patching workload expressed in number of vulnerabilities to fix identified in NVD. The ordering of the risk reduction measure is essentially preserved over all software categories with the exception of \WIN, for which additional considerations apart from the technical characteristics of the vulnerability may need to be considered.
\vspace{0.05in}
\end{minipage}

\begin{tabular}{p{3.5cm}rrrrrrrrrr}
&\multicolumn{2}{c}{Aggregate}&\multicolumn{2}{c}{\IE}&\multicolumn{2}{c}{\PLU}&\multicolumn{2}{c}{\WIN}&\multicolumn{2}{c}{\COM}\\
\toprule
Policy&RR&\#V&RR&\#V&RR&\#V&RR&\#V&RR&\#V\\\midrule
$All\ vulns$& - &14380&-&1223&-&588&-&490&-&895\\
$ CVSS\ \geq 4$&23.2\%&13715&-24.4\%&1208&50.5\%&577&3.4\%&477&23.1\%&867\\
%$CVSS \geq 9$&8.9\%&3081&-9.2\%&452&-0.3\%&368&21.9\%&174&-11.3\%&639\\
$ Comp=L$&13\%&7552&7.9\%&635&5.2\%&230&-5.4\%&236&11.4\%&186\\
$ Comp \leq M$&5.2\%&13952&-37.4\%&1197&36.4\%&561&-23.0\%&472&-15.5\%&878\\
$ E[pA]=H$&26.8\%&8308&-28.2\%&733&52.2\%&449&-52.4\%&389&21.3\%&739\\\bottomrule
\end{tabular}
\end{table*}
As we can see from the table, the relative ordering is essentially preserved for all control factors. With few exceptions, the preferred global policy remains the preferred one even when restricted to a specific 
software category. \WIN\ vulnerabilities are here an exception, showing
that the trade-off may not be a good proxy for this software category and that other variables should be considered. \IE\ shows mixed behavior with $ E[pA]=H$ performing similarly to $ CVSS\ \geq 4$, but worse than $Comp=L$ in terms of $RR$. A negative $RR$ indicates that the residual risk in the `untreated' vulnerabilities is \emph{higher} than the risk for the treated vulnerabilities.

% Results in Table \ref{tab:results} show that the impact-complexity trade-off
% can be used a basis to design more efficient patching policies.
% However, the mere existence of an exploit does not necessarily reflect
% the real `risk in the wild' posed by the vulnerability: previous
% research showed that certain vulnerabilities are addressed
% more often by attackers than others, and should therefore
% be prioritised in terms of fixing procedures \cite{Allodi-ESSOS-15,Nayak-2014-RAID}. It is therefore insightful to evaluate the effectiveness
% of each policy with respect to the actual reduction in terms
% of $pA$ the policy implies.
% We will now validate whether our policy prioritization is correct by introducing the notion of \emph{Potential of Attack}.

\subsection{$pA$ reduction in the wild}

% Obviously the actual level of risk reduction changes when we consider the individual controls.

% For example, for \emph{PoC}-based policies, the additional condition of selecting vulnerabilities with a high CVSS brings the highest RR for all controls excepts for WINDOWS. For the latter case however, all RRs are flattened in a 3-4\% range so this difference may not be significant.

% For \emph{Exist}-base policies there is one of the exceptions to the preferred ordering. The \IE\ category is the significant exception for the existential base policies. For browsers the mere existence of a vulnerability is not a sufficient conditions, even if this condition is refined by a technical analysis of the vulnerability.
% In this category the $Comp=L$ is the preferred policy with the best risk reduction (a positive albeit small one) all others are negative, i.e. not effective. 

We now look at the the efficacy of each policy in reducing attacks in the wild. Table~\ref{tab:results_pA}
\begin{table*}[t]
\caption{Workload and $pA$ reduction for all controls.}\label{tab:results_pA}
\centering
\small
\begin{minipage}{0.9\textwidth}
\footnotesize
The policy based on the $E[pA]$ measure drastically decreases the fraction of vulnerabilities to consider for patching while foiling the vast majority of attacks in the wild (6.7 $pA$ points against an overall $pA$ of 6.8).
\vspace{0.05in}
\end{minipage}
\begin{tabular}{  p{3.5cm}  r r  r r  r r  r r  r r }
\toprule
&\multicolumn{2}{c}{Aggregate}&\multicolumn{2}{c}{\IE}&\multicolumn{2}{c}{\PLU}&\multicolumn{2}{c}{\WIN}&\multicolumn{2}{c}{\COM}\\%\cline{2-11}
Policy&\%V&$pA$&\%V&$pA$&\%V&$pA$&\%V&$pA$&\%V&$pA$\\\midrule
$All\ vulns$&100.0\%&6.8&100.0\%&5&100.00\%&6.7&100.0\%&6.1&100.0\%&4.4\\
$ CVSS\ \geq 4$&95.4\%&6.8&98.8\%&5&98.13\%&6.7&97.3\%&6.1&96.9\%&4.4\\
%$CVSS \geq 9$&21.4\%&6.7&37.0\%&5&62.59\%&6.5&35.5\%&6.1&71.4\%&4.4\\
$Comp=L$&52.5\%&6.3&51.9\%&-&39.12\%&6.3&48.2\%&5&20.8\%&3.3\\
$Comp \leq M$&97.0\%&6.8&97.9\%&5&95.41\%&6.7&96.3\%&6.1&98.1\%&4.4\\
$E[pA]=H$&57.8\%&6.7&59.9\%&5&76.36\%&6.6&79.4\%&6.1&82.6\%&4.4\\\bottomrule
% \multirow{6}{*}{$PoC$}&None&21.1\%&6.5&4.2\%&5&7.48\%&6.3&16.9\%&6.1&14.7\%&4.2\\
% &$PoC+ CVSS \geq 4$&20.9\%&6.5&4.1\%&5&7.48\%&6.3&16.7\%&6.1&14.5\%&4.2\\
% &$PoC+ CVSS \geq 9$&3.8\%&6.5&2.2\%&5&6.12\%&6.3&8.6\%&6.1&9.3\%&4.2\\
% &$PoC+ Comp=L$&13.7\%&5.6&1.4\%&-&3.40\%&5.6&6.1\%&4.7&3.6\%&3\\
% &$PoC+ Comp \leq M$&20.7\%&6.5&4.0\%&5&7.31\%&6.3&15.9\%&6.1&14.6\%&4.2\\
% &$PoC+ Comp \leq M \& Imp \geq M$&16.7\%&6.5&2.5\%&5&7.14\%&6.3&14.9\%&6.1&12.4\%&4.2\\\hline\hline
% \multirow{6}{*}{$Black$}&None&0.4\%&6.3&1.0\%&4.8&5.44\%&6.3&1.2\%&4.8&0.3\%&3.1\\
% &$Black+ CVSS \geq 4$&0.4\%&6.3&1.0\%&4.8&5.44\%&6.3&1.2\%&4.8&0.3\%&3.1\\
% &$Black+ CVSS \geq 9$&0.3\%&6.3&0.8\%&4.8&4.93\%&6.3&1.0\%&4.8&0.3\%&3.1\\
% &$Black+ Comp=L$&0.1\%&5.5&0.2\%&-&2.55\%&5.5&0.0\%&-&0.1\%&3\\
% &$Black+ Comp \leq M$&0.4\%&6.3&0.9\%&4.8&5.44\%&6.3&1.0\%&4.8&0.3\%&3.1\\
% &$Black+ Comp \leq M \& Imp \geq M$&0.4\%&6.3&0.8\%&4.8&5.44\%&6.3&1.0\%&4.8&0.3\%&3.1\\\hline\hline
\end{tabular}
\end{table*}
reports, for each control, the amount of patching work required relative to
the total for that software category and the reduction in $pA$.
% At first glance we see that most of the  $pA$ columns have the same value across the rows. It should be noticed that the actual values are not equal. $pA$ is a logarithm in base 10 and it is truncated to the first decimal. It offers a bird's eye view on the attacks, eliminating most of the noise. It shows that if we consider the order of magnitudes of attacks there is very little difference between `high workload' policies and `low workload' ones.
% In particular, 
By looking at the $pA$ and $\%V$ columns,
it is possible to see that the $E[pA]=H$
patching policy achieves the lowest workload (requiring
to patch only 58\% of the original volume of vulnerabilities),
and still fully addresses $pA$ in the wild
throughout all software categories\footnote{No value is reported 
in Table \ref{tab:results_pA} for
$Comp = L$ under the \IE\ category because there are no
attacks of this type in this category.}.

Further, the results reported in Table \ref{tab:results_pA}
can be used to validate the risk reduction measure
introduced in \cite{Allodi-2014-TISSEC}. For RR
to be a valid measure for patching policy effectiveness, 
there should be a correspondence between the level of RR and 
the policy's effectiveness in the wild.
In Table \ref{tab:full_results} the `PCI DSS' policy (patch all vulnerabilities with $CVSS\geq4$) and $E[pA]=H$
have the highest risk reductions (23\% and 26\% respectively).
$Compl=L$ and $Comp \leq M$ showed a much lower RR, the latter
being the worst. We find this same ordering to be preserved in the 
evaluation in the wild reported in Table \ref{tab:results_pA}.
$E[pA]=H$ represents the best
trade-off between workload and reduction in $pA$, while
$Comp=L$ entails the lowest workload but at the price of a lower
reduction in $pA$.
Similar considerations can be done if we break down the analysis for the 
controls.
%  Hence, using risk reduction to select a rule-based policy 
% to be a very effective strategy in terms of existence of attacks and also in 
% terms or reduction of attack potentials i.e. the magnitude of the number of 
% attacks.

% \subsection{Implementation in the CVSS v3 standard.} Two of the authors
% of the present paper worked with the First.org SIG team to define
% the new upcoming version of the CVSS standard (v3). This work in particular
% has been instrumental to help define the Attack Complexity metric in the new
% version of the standard.
% Vulnerability assessments done with the old v2 definition of \texttt{Access Complexity} 
% are able to capture an important phenomenon: the level of \texttt{Access Complexity} 
% is a good predictor of the expected volume of attacks against that 
% vulnerability in the wild. Attacks against Low complexity vulnerabilities 
% would be listed into the millions while High complexity vulnerabilities would 
% only score hundreds.
% This can be particularly useful in practice when the CVSS assessment is used 
% to prioritize patching work and/or assess vulnerability risk (e.g. notably: 
% NIST SCAP protocol, PCI DSS). This effect is lost with the current v3 
% definition of AC. An example is reported in Appendix.
% UNITN investigated the effect on a sample of vulnerabilities and found that 
% the missing bit can be traced back to a fragment of v2’s definition: “Some 
% information must be gathered before a successful attack can be launched.” [
% CVSS v2 Guide, \texttt{Access Complexity}]. 

\section{Threats to validity}\label{sec:threats}

% Because of the data collection methodology, the 
% validity of our conclusions may be
% limited by a number of factors.
\emph{External validity}. \WINE\ data is a representative sample of
attacks  detected in the wild  against `user machines'.
Our conclusions are therefore
limited to systems of the same nature: different dynamics may hold for server machines or specialized systems.
%  We refrain from
% extending our results to server machines or critical systems
% subject to targeted attacks, for which different
% attack dynamics may hold.
\emph{Internal validity}. Volumes of attacks against
vulnerabilities may 
change by geographical area. 
It is possible that some vulnerabilities attacked only 
in particular areas or affecting only particular systems of lower commercial 
interest for Symantec
may not appear or are 
under-represented in our datasets. To address this, in our case-control
study we control for possible factors for inclusion in \SYM\ \cite{Allodi-2014-TISSEC}.
Further 
refinements %in population control 
may 
be needed to safely narrow the scope of our 
conclusions
down to specific user populations.

% Because
% some attack signatures refer $n>1$ exploited vulnerability,
% we normalise by $n$. This maintains the desirable property of not creating
% attacks that do not exist (by summing up the same attack $n$ times). On the 
% other hand it may introduce noise in the data, as some
% CVEs may be \emph{under} estimated in terms of delivered
% attacks. However, because only 3\% of signatures in \WINE\ report
% 3 or more CVEs in \SYM, the impact in terms of noise is minimal. 

% Moreover, it must be underlined that in this
% paper we refer only to \emph{network} attacks (e.g. web delivered attacks).
% We do not consider here vulnerabilities exploited by viruses and malware in general.

%It is possible that our four datasets do not correctly capture the real
%distribution of CIA assessments.
%We don't identify any threat to the validity of our conclusions for
%the \NVD\ and \EDB\ datasets: these are not sampled,
%and are the reference datasets for vulnerabilities
%and proof-of-concept exploits respectively.
%
%\EKITS, on the other hand, is the dataset of vulnerabilities featured in
%exploit kits.
%Because of the peculiar attack dynamics of these tools,
% \EKITS\ may be not representative of the whole population
%of vulnerabilities traded in the black markets, but solely of those featured in
%Exploit Kits.
%
%Similarly, \SYM\ features vulnerabilities detected by Symantec's
%products, that may be of a particular type only (that of interest for Symantec).
%However, all the 9
%meaningful CIA combinations in \NVD\ are featured in \SYM\ as well.

\section{Related work} \label{sec:related}

An analysis of the distribution of
CVSS scores and subscores has been presented by Scarfone et al. in
\cite{SCAR-MELL-09-ESEM}.
Frei et al.'s \cite{Frei-2006-LSAD} studied the life-cycle of a
vulnerability from exploit to patch. Their dataset is a composition of \NVD,
OSVDB and `FVDB' (Frei's Vulnerability DataBase, obtained from the examination
of security advisories for patches). The notion of vulnerability risk
has been considered in \cite{Allodi-2014-TISSEC}, where the authors
show that the CVSS score as an aggregate number 
is not a satisfactory risk metric for 
vulnerabilities.  Holm et al. \cite{holm2015expert} investigated through 
expert opinion 
which aspects of a vulnerability should be considered on top of the baseline 
CVSS assessment to better represent risk of exploit. 
%Other authors employed the bare \texttt{CVSS-score} as a metric for vulnerability risk~\cite{Naaliel-ISSRE-14,Houmb-2010-JSS}.
The present work extends this line of research by
showing that the inner assessments in the CVSS framework can be instrumental
for vulnerability prioritisation. Other work focused on the volume
of attacks in the wild recorded against vulnerabilities. Allodi showed
in \cite{Allodi-ESSOS-15} that vulnerability exploitation follows
a heavy-tailed distribution, and that for some software types
as little as 5\% of attacked vulnerabilities represent 95\% of
the risk in the wild. Similarly, Nayak et al. \cite{Nayak-2014-RAID} showed
that attackers prefer certain vulnerabilities over others. Our work
integrates these results by showing that part
of the attacker's decision process is influenced by a trade-off
between the 
complexity and the impact of the vulnerability exploit.

%%%%%%%%%%%%%%%%%%%%%%%%%%%%%%%%%%%%%%%%%%%%%%%%%%%%%%%%%%%%%%%%%%%%%%%%%%%%%%%%%%
\section{Conclusions}\label{sec:conclusions}

In this paper we investigated trends in vulnerability attacks per
CVSS Impact and Exploitability types. We find that %:
% \begin{enumerate}
% \item The CIA Impact assessment can be, as-is, more noisy than
% informative: three CIA Impact types are enough to represent the overwhelming
% majority of attacks (\sf{CC*},\sf{PP*},\sf{**P}). Simplifying the assessment to solely three categories,
% namely \textsf{Total Compromise, Partial Compromise, Crash} can result
% in an easier to manage and more informative metric without losing expressive
% detail.
%\item 
there exists a clear-cut distinction in terms
of attacks in the wild between low complexity, high impact vulnerabilities and high complexity, low impact vulnerabilities.
% between vulnerability impact and
% exploitation complexity. The attackers aim with much higher intensity
% at either easy to exploit vulnerabilities, or more difficult vulnerabilities
% provided that their impact is maximum.
%\end{enumerate}

% \section{Discussion}
% \label{sec:discussion}

% In this paper we show that part of the attacker
% decision process in choosing which vulnerabilities to attack
% \emph{en-masse} is ruled by a simple trade-off mechanism
% whereby vulnerabilities with a high impact and low exploitation complexity
% are preferred by the attacker.

% This is reflected in both the number of exploited vulnerabilities
% and in the volume of attacks
% received for each \emph{Impact,Complexity} tuple.
% We leverage this effect to build an estimator that, given the
% CVSS assessment of the vulnerability, can discern between vulnerabilities
% with a high potential of attack and vulnerabilities with a low potential of
% attack. This can be particularly useful in practice to prioritise 
% patching work, as vulnerabilities with a high attack potential
% represent higher risk and therefore should be fixed with higher priority.
% The estimator presented in this work is straightforward to implement
% and can be used in foresight without pre-existing data on the
% volume of attacks in the wild for that vulnerability.
Leveraging this effect we build an estimator of the \emph{Attack Potential} of a vulnerability that provides a first indicator of the risk of exploitation in the wild. 
The estimator presented in this work is straightforward to implement
and can be used in foresight without pre-existing data on the
 volume of attacks in the wild for that vulnerability.
% This can be particularly useful in practice to prioritise 
% patching work, as vulnerabilities with a high attack potential
% represent higher risk and therefore should be fixed with higher priority.
We test our estimator against standard CVSS-based patching policies and show that it outperforms them in terms of foiled attacks and entailed workload. The Attack Complexity metric in the 3.0 release of CVSS reflects these observations.
% This also serves as a validation of the Risk Reduction metric as a useful
% measure to estimate policy effectiveness.

% Two of the authors (Allodi, Massacci)
% of the present paper worked with the First.org SIG team to define
% the new upcoming version of the CVSS standard (v3). This work in particular
% has been instrumental to help define the Attack Complexity metric in the new
% version of the standard.

\section*{Acknowledgements}
 % This project has received funding from the European Union's Seventh Framework Programme under grant agreement no 285223 (SECONOMICS). 
 This work has been supported by the EU FP7 programme (grant no. 285223, SECONOMICS) and by the NWO  project SpySpot (grant no. 628.001.004).
% %We also would like to thank Jeffrey Wilhelm
% %of Symantec Research Labs for the useful feedback.
 Interested researchers can reproduce our results by accessing
 the WINE dataset 2012-008.

\bibliographystyle{abbrv}
\bibliography{short-names,security-common}

\begin{thebibliography}{10}

\bibitem{Allodi-ESSOS-15}
L.~Allodi.
\newblock The heavy tails of vulnerability exploitation.
\newblock In {\em Proc.\ of ESSoS'15}, 2015.

\bibitem{Allodi-2014-TISSEC}
L.~Allodi and F.~Massacci.
\newblock Comparing vulnerability severity and exploits using case-control
  studies.
\newblock {\em ACM Transaction on Information and System Security (TISSEC)},
  17(1), August 2014.

\bibitem{Allodi-17-WAAM}
L.~Allodi, F.~Massacci, and J.~Williams.
\newblock The work-averse cyber attacker model. evidence from two million
  attack signatures.
\newblock In {\em Published in WEIS 2017. Available at
  \url{https://ssrn.com/abstract=2862299}}, 2017.

\bibitem{Allodi-2013-IWCC}
L.~Allodi, S.~Woohyun, and F.~Massacci.
\newblock Quantitative assessment of risk reduction with cybercrime black
  market monitoring.
\newblock In {\em In Proc.\ of IWCC'13}, 2013.

\bibitem{chakradeo2013mast}
S.~Chakradeo, B.~Reaves, P.~Traynor, and W.~Enck.
\newblock Mast: Triage for market-scale mobile malware analysis.
\newblock In {\em Proceedings of the sixth ACM conference on Security and
  privacy in wireless and mobile networks}, pages 13--24. ACM, 2013.

\bibitem{Christey-2013-BHUSA}
S.~Christey and B.~Martin.
\newblock Buying into the bias: why vulnerability statistics suck.
\newblock \url{https://www.blackhat.com/us-13/archives.html\#Martin}, July
  2013.

\bibitem{Dumitras-11-BADGERS}
T.~Dumitras and D.~Shou.
\newblock Toward a standard benchmark for computer security research: The
  worldwide intelligence network environment (wine).
\newblock In {\em Proc.\ of BADEGRS'11}, pages 89--96. ACM, 2011.

\bibitem{Evans-1986-ACC}
L.~Evans.
\newblock The effectiveness of safety belts in preventing fatalities.
\newblock {\em Accident\ Anal.\ \& Prev.}, 18(3):229--241, 1986.

\bibitem{Frei-2006-LSAD}
S.~Frei, M.~May, U.~Fiedler, and B.~Plattner.
\newblock Large-scale vulnerability analysis.
\newblock In {\em Proc.\ of LSAD'06}, pages 131--138. ACM, 2006.

\bibitem{holm2015expert}
H.~Holm and K.~K. Afridi.
\newblock An expert-based investigation of the common vulnerability scoring
  system.
\newblock {\em Computers \& Security}, 53:18--30, 2015.

\bibitem{Mell-2007-CMU}
P.~Mell, K.~Scarfone, and S.~Romanosky.
\newblock A complete guide to the common vulnerability scoring system version
  2.0.
\newblock Technical report, FIRST, Available at
  \url{http://www.first.org/cvss}, 2007.

\bibitem{Nayak-2014-RAID}
K.~Nayak, D.~Marino, P.~Efstathopoulos, and T.~Dumitra{\c{s}}.
\newblock Some vulnerabilities are different than others.
\newblock In {\em Proc.\ of RAID'14}, pages 426--446. Springer, 2014.

\bibitem{PCI-DSS-DOC}
PCI.
\newblock Pci dss requirements and security assessment procedures, version 2.0.
\newblock \url{https://www.pcisecuritystandards.org/documents/pci_dss_v2.pdf},
  2010.

\bibitem{Scarfone-2010-SCAP}
S.~D. Quinn, K.~A. Scarfone, M.~Barrett, and C.~S. Johnson.
\newblock Sp 800-117. guide to adopting and using the security content
  automation protocol (scap) version 1.0.
\newblock Technical report, NIST, 2010.

\bibitem{SCAR-MELL-09-ESEM}
K.~Scarfone and P.~Mell.
\newblock An analysis of cvss version 2 vulnerability scoring.
\newblock In {\em Proc.\ of ESEM'09}, pages 516--525, 2009.

\bibitem{Verizon-2014}
Verizon.
\newblock Verizon 2014 pci compliance report.
\newblock Technical report, Verizon Enterprise, 2014.

\end{thebibliography}

\end{document}